\documentclass[twocolumn,tighten,twocolappendix]{aastex63}
\pdfoutput=1 
\usepackage{amsmath,amstext}
\usepackage[T1]{fontenc}
\usepackage{ae,aecompl}
\usepackage[utf8]{inputenc}
\usepackage{apjfonts} 
\usepackage[figure,figure*]{hypcap}
\usepackage{enumitem}
\usepackage{bm}
\usepackage{graphicx}
\input{hyperlink-year-only-natbib-patch}

\newcommand{\pc}{\ensuremath{{\rm pc}}}
\newcommand{\kpc}{\ensuremath{{\rm kpc}}}
\newcommand{\mpc}{\ensuremath{{\rm Mpc}}}
\newcommand{\kms}{\ensuremath{{\rm km\,s^{-1}}}}

\newcommand{\Gyr}{\ensuremath{{\rm Gyr}}}

\newcommand{\K}{\ensuremath{{\rm K}}}
\newcommand{\s}{\ensuremath{{\rm s}}}

\newcommand{\msun}{\ensuremath{{M_{\odot}}}}

\shortauthors{Nadler, Yang, \& Yu}

\begin{document}

\title{A Self-interacting Dark Matter Solution to the Extreme Diversity of Low-mass Halo Properties}
\shorttitle{Extreme Halo Diversity in SIDM}

\author[0000-0002-1182-3825]{Ethan O.~Nadler}
\affiliation{Carnegie Observatories, 813 Santa Barbara Street, Pasadena, CA 91101, USA}
\affiliation{Department of Physics $\&$ Astronomy, University of Southern California, Los Angeles, CA 90007, USA}

\author[0000-0002-5421-3138]{Daneng Yang}
\affiliation{Department of Physics and Astronomy, University of California, Riverside, CA 92521, USA}

\author[0000-0002-8421-8597]{Hai-Bo Yu}
\affiliation{Department of Physics and Astronomy, University of California, Riverside, CA 92521, USA}

\email{enadler@carnegiescience.edu}
\email{danengy@ucr.edu}
\email{haiboyu@ucr.edu}

\begin{abstract}
The properties of low-mass dark matter (DM) halos appear to be remarkably diverse relative to cold, collisionless DM predictions, even in the presence of baryons. We show that self-interacting DM (SIDM) can simultaneously explain observations of halo diversity at two opposite extremes---the inner density profile of the dense substructure perturbing the strong lens galaxy SDSSJ0946+1006 and the rotation curves of isolated, gas-rich ultradiffuse galaxies (UDGs). To achieve this, we present the first cosmological zoom-in simulation featuring strong DM self-interactions in a galaxy group environment centered on a $10^{13}~\msun$ host halo. In our SIDM simulation, most surviving subhalos of the group-mass host are deeply core-collapsed, yielding excellent candidates for the observed dense strong-lens perturber. Self-interactions simultaneously create kiloparsec-scale cores in low-concentration isolated halos, which could host the observed UDGs. Our scenario can be further tested with observations of DM structure and galaxies over a wide mass range.
\end{abstract}

\keywords{\href{http://astrothesaurus.org/uat/353}{Dark matter (353)};
\href{http://astrothesaurus.org/uat/261}{Strong gravitational lensing (261)};
\href{http://astrothesaurus.org/uat/940}{Low surface brightness galaxies (940)};
\href{http://astrothesaurus.org/uat/1083}{N-body simulations (1083)};
\href{http://astrothesaurus.org/uat/1880}{Galaxy dark matter halos (1880)}}

\section{Introduction} The dark matter (DM) distribution on galactic scales and below is a powerful probe of DM particle properties (e.g., \citealt{BechtolSnowmass}). One key measure of this distribution is its \emph{diversity}, i.e., the range of DM halo properties at fixed mass. Studies have long shown that the inner circular velocities of spiral and dwarf irregular galaxies in the field span a wide range---in excess of cold, collisionless DM (CDM)-only predictions---even though their total halo masses are similar~\citep{deNaray09123518,Oman150401437,Ren:2018jpt,Santos-Santos191109116}. Although baryonic feedback within galaxies likely explains some of this diversity (e.g., see \citealt{Sales220605295} for a review), more recent observations challenge this as a complete explanation.

In particular, \defcitealias{Minor201110627}{M21} Minor et al.\ (\citeyear{Minor201110627}, hereafter \citetalias{Minor201110627}) showed that the strong lens SDSSJ0946+1006 is perturbed by a DM (sub)structure~\citep{Vegetti09100760} with an extremely high density, $\gtrsim3\sigma$ ($\gtrsim5\sigma$) away from the median CDM expectation when interpreted as a subhalo (line-of-sight halo), and \cite{Ballard230904535} reported a comparable (though slightly weaker) tension. Meanwhile, a class of large, gas-rich isolated galaxies has recently been discovered~\citep{Leisman170305293,ManceraPina:2019zih,ManceraPina:2020ujo,PinaMancera211200017,guo190800046,Shi:2021tyg}. These ultradiffuse galaxies (UDGs) contain less DM than expected based on their baryonic content; \defcitealias{Kong220405981}{K22} Kong et al.~(\citeyear{Kong220405981}, hereafter \citetalias{Kong220405981}) showed that these UDGs must inhabit extremely low-concentration halos, $\gtrsim 3\sigma$ to $5\sigma$ away from the median CDM expectation (also see~\citealt{ManceraPina:2019zih,PinaMancera211200017}). However, \citetalias{Kong220405981} find that CDM halos from the IllustrisTNG hydrodynamic simulation~\citep{Nelson181205609}, even with such low concentrations, are still much denser than these UDGs.

Baryonic feedback can steepen (sub)halo density profiles, via adiabatic contraction \citep{Blumenthal1986,Gnedin0406247}, or core them via repeated cycles of supernova feedback \citep{Navarro9610187,Governato:2009bg,Pontzen11060499}. 
However, it is challenging for baryons to simultaneously explain the diverse observations described above. In particular, the SDSSJ0946+1006 perturber does not have a detectable luminous component; thus, its stellar mass is negligible and that adiabatic contraction is unlikely to explain its extremely steep density profile~(\citealt{Vegetti09100760}; \citetalias{Minor201110627}). If baryonic feedback forms a core in the progenitor halo, the tension would be exacerbated.

At the opposite extreme, gas-rich UDGs have very shallow potentials; thus, it takes longer to replenish the gas necessary for bursty star formation. As a result, supernova feedback is expected to be inefficient in UDGs~(\citealt{PinaMancera211200017}; \citetalias{Kong220405981}), although we note that some hydrodynamic simulations with episodic feedback produce diffuse galaxies in median-concentration halos (e.g., \citealt{DiCintio:2016ehs,Chan171104788}). In addition, environmental effects such as tidal stripping---while important for UDGs in groups and clusters~\citep{Jiang:2018iut,Ogiya:2018jww,Carleton180506896,Tremmel190805684,Sales:2019iwl,Yang200202102,Benavides210901677,Moreno:2022srs}---are largely absent for the UDGs analyzed in \citetalias{Kong220405981}.

We will show that both tensions may be resolved if DM has strong self-interactions that thermalize halos' inner regions~(see~\citealt{Tulin170502358,Adhikari220710638} for reviews). The key physics that enables this is the gravothermal evolution of self-interacting DM (SIDM) halos, which have two sequential phases: core expansion and collapse. In the first phase, interactions transport heat from outer to inner halo regions, lowering the central density; in the second, the direction of heat flow reverses and the inner halo becomes denser than its CDM counterpart~\citep{Balberg0110561,Koda11013097}. The timescale for the onset of core collapse scales as $t_c\propto (\sigma_{\rm eff}/m)^{-1}c^{-7/2}_{200}M^{-1/3}_{200}$~\citep{Essig180901144,Kaplinghat190404939}, where $\sigma_{\rm eff}/m$ is an effective cross section~\citep{Yang220503392,Outmezguine220406568,Yang220502957} and $M_{200}$ ($c_{200}$) are mass (concentration), respectively. If $t_c\sim{\cal O}(10)~{\rm Gyr}$, the diversity of inner halo densities encoded by concentration~\citep{Kaplinghat150803339,Kamada161102716} is amplified at~$z\sim0$~\citep{Yang221113768}.

In this Letter, we introduce a new velocity-dependent SIDM model, perform the first high-resolution cosmological $N$-body simulation with strong DM self-interactions on galaxy group scales, and identify simulated halos and subhalos that can self-consistently explain the observed lensing perturber and UDGs.

\section{SIDM Model} We consider velocity-dependent DM self-interactions with a Rutherford-like differential cross section~\citep{Feng09110422,Ibe:2009mk,Tulin13023898,Yang220503392}
\begin{eqnarray}
\label{eq:xsr}
\frac{\mathrm{d}\sigma}{\mathrm{d} \cos\theta} = \frac{\sigma_{0}w^4}{2\left[w^2+{v^{2}}\sin^2(\theta/2)\right]^2 },  
\end{eqnarray}
where $\theta$ is the scattering angle in the center of mass frame, $v$ is the relative velocity of DM particles, the cross section drops as $v^{-4}$ for $v>w$, and $\sigma_0$ controls the normalization. We set the cross section per mass of DM particles $\sigma_0/m=147.1~\rm cm^2~g^{-1}$, $w=120~\rm km~s^{-1}$ and refer to this as the ``Group SIDM'' model. A cross section with the same amplitude but a lower turnover scale of $w=24.33~\rm km~s^{-1}$ was simulated in a Milky Way (MW) setting (\citealt{Yang221113768}; also see \citealt{Zavala190409998,Correa200702958,Turner201002924,Correa220611298}). 

In general, DM scattering is both velocity- and angle-dependent, as indicated in Equation~\ref{eq:xsr}.~\cite{Yang220503392} demonstrated that the angular dependence can be absorbed by introducing a viscosity cross section, which weights the differential cross section with a factor of $\sin^2\theta$. Figure~\ref{fig:bmxs} (inset) shows the viscosity cross section $\sigma_{V}$ for the SIDM model we consider; see Appendix~\ref{sec:cross_section} for details. Since $\sigma_V$ does not explicitly depend on the scattering angle, it greatly simplifies the implementation of DM self-interactions in $N$-body simulations. Our SIDM simulation will be based on the velocity-dependent viscosity cross section.
 
For a given halo, its gravothermal evolution can be captured by the constant effective cross section~\citep{Yang220503392}
\begin{equation}
\label{eq:eff}
\sigma_{\rm eff} = \frac{2 \int v^2\mathrm{d} v \mathrm{d} \cos\theta \frac{\mathrm{d} \sigma}{\mathrm{d} \cos\theta} \sin^2\theta v^5 \exp \left[-\frac{v^2}{4\nu_{\rm eff}^2}\right]}{\int v^2\mathrm{d} v \mathrm{d} \cos\theta \sin^2\theta v^5  \exp \left[-\frac{v^2}{4\nu_{\rm eff}^2}\right]}, 
\end{equation}
where $\nu_{\rm eff}\approx0.64V_{\rm max}$ is a characteristic velocity dispersion for a CDM halo with maximum circular velocity $V_{\rm max}$. Figure~\ref{fig:bmxs} shows $\sigma_{\rm eff}/m$ versus $V_{\rm max}$ for our Group SIDM model (red), the model in~Yang et al.~(\citeyear{Yang221113768}, purple), which is similar to that in~\cite{Turner201002924}, and the best-fit model from Kaplinghat et al.~(\citeyear{Kaplinghat150803339}, black) for comparison. The effective cross section of our SIDM model can be fitted with a functional form of $147.1~\mathrm{cm^2~g}^{-1}/[1+(V_{\rm max}/80~{\rm km~s^{-1}})^{1.72}]^{1.90}$. For a Navarro--Frenk--White (NFW) halo~\citep{Navarro:1995iw}, $V_{\rm max}=1.64 r_s\sqrt{G\rho_s}$, where $\rho_s$ is its scale density and $r_s$ its scale radius, and $G$ is Newton's constant. In this case, the effective cross section is ultimately related to the halo's mass $M_{200}$ and concentration $c_{200}$. For a halo with a given $V_{\rm max}$, the effective cross section provides a good proxy to evaluate the impact of the self-interactions on halo properties, although the scattering probability of individual DM particles is velocity- and angle-dependent; see~\cite{Yang220503392} for further details.

\begin{figure}[t!]
  \centering
  \includegraphics[height=8.2cm]{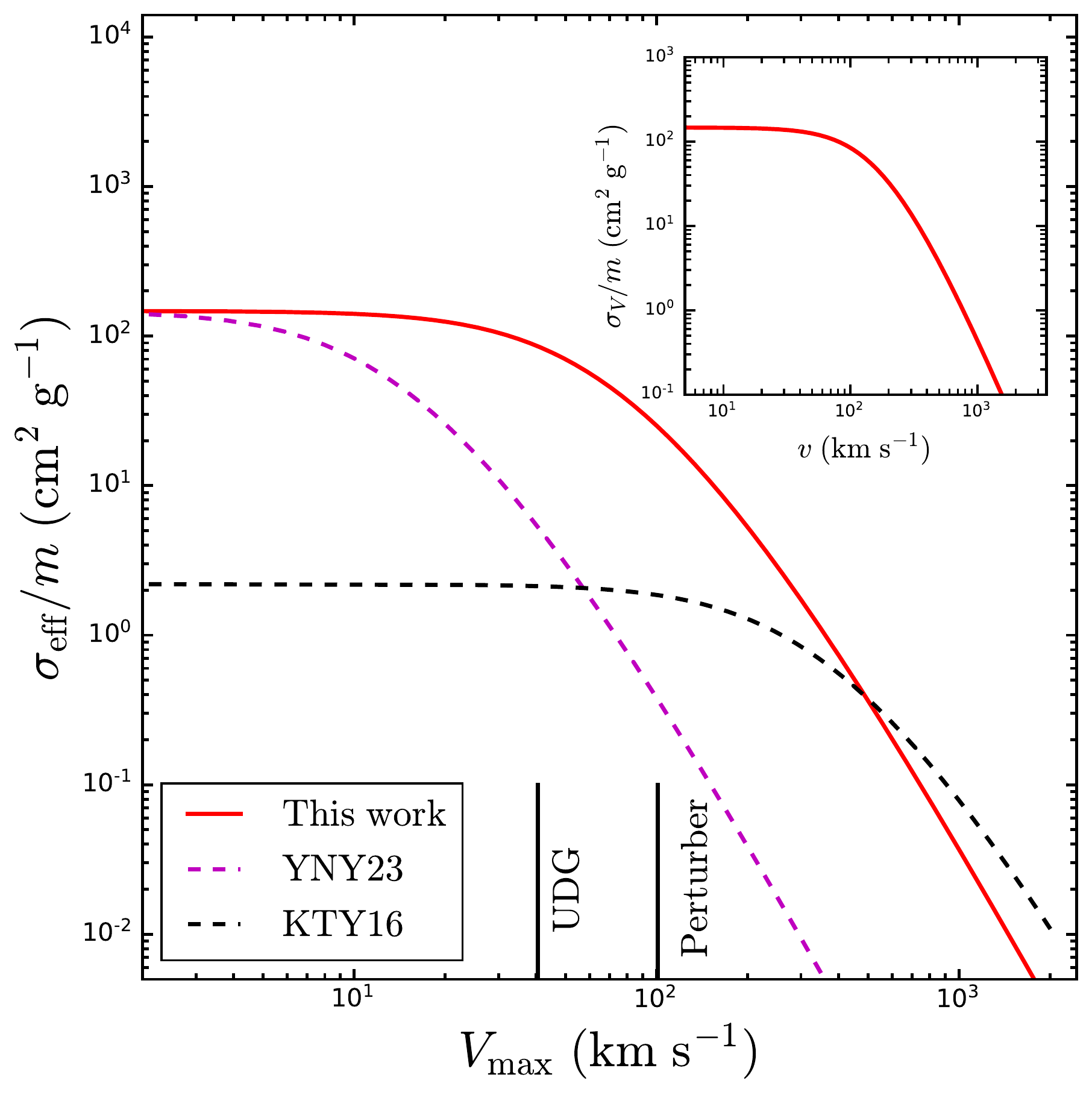}
  \caption{\label{fig:bmxs} The effective DM self-interaction cross section (main panel) and viscosity cross section (inset) for our Group SIDM model (red), and for the SIDM models from Yang et al.~(\citeyear{Yang221113768}, purple) and Kaplinghat et al.~(\citeyear{Kaplinghat150803339}, black). Vertical dotted lines show the approximate velocity scales of the UDGs and strong-lens perturber.}
\end{figure}

Our Group SIDM model is motivated as follows. First, a cross section significantly larger than $\sim1~{\rm cm^2~g^{-1}}$ is needed to produce a shallow density profile in low-concentration halos (e.g., see ~\citetalias{Kong220405981}). For a typical UDG halo with $V_{\rm max}=40~{\rm km~s^{-1}}$ and $c_{200}$ $5\sigma$ below the CDM median, our Group SIDM model predicts $\sigma_{\rm eff}/m\approx86~{\rm km~s^{-1}}$, and the collision rate at the scale radius is $1.3$ per $10~{\rm Gyr}$, high enough to produce a density core. Second, for (sub)halos with $V_{\rm max}\approx100~{\rm km~s^{-1}}$, relevant for the lensing perturber, $\sigma_{\rm eff}/m\approx25~{\rm cm^2~g^{-1}}$ and $t_c\lesssim10~{\rm Gyr}$ if $c_{200}$ is $1.5\sigma$ higher than the CDM median, where we have used the following formula to estimate the collapse time ~\citep{Essig180901144} 
\begin{eqnarray}
    t_c=\frac{150}{C}\frac{1}{r_s\rho_s(\sigma_{\rm eff}/m)}\frac{1}{\sqrt{4\pi G \rho_s}},
\end{eqnarray}
with $C=0.75$ from calibrating against $N$-body simulations. Further using the relations
\begin{equation}
r_s=\left[\frac{3 M_{200}}{(200\rho_{\rm crit})4\pi c^3_{200}}\right]^{\frac{1}{3}},~\rho_s=\frac{200\rho_{\rm crit}c^3_{200}}{3f(c_{200})},
\end{equation}
where $\rho_{\rm crit}$ is the critical density and $f(c_{200})\equiv\ln(c_{200}+1)-c_{200}/(c_{200}+1)$, we have $t_c\propto (\sigma_{\rm eff}/m)^{-1}c^{-7/2}_{200}M^{-1/3}_{200}$ as $f$ has a very mild dependence on $c_{200}$~\citep{Essig180901144}. Thus, strong DM self-interactions can further diversify inner halo densities encoded by concentration; for subhalos, tidal stripping can accelerate core collapse~\citep{Nishikawa190100499,Sameie190407872,Kahlhoefer190410539,Correa200702958,Zeng211000259}.

Finally, our model evades constraints of $\sigma_{\rm eff}/m\lesssim1~$~\citep{Sagunski:2020spe} and $0.1~{\rm cm^2~g^{-1}}$~\citep{Kaplinghat150803339,Andrade2020} from groups and clusters ($V_{\rm max}\approx800$ and $1000~{\rm km~s^{-1}}$, respectively). Cross section constraints from cluster mergers are typically weaker~\citep{Harvey:2015hha,Jauzac:2016tjc,Wittman:2017gxn}.

We also comment on constraints from MW satellites and spiral galaxies in the field. Previous studies have shown that MW satellites' DM distributions are diverse, and that it is difficult to accommodate the full diversity of these systems in SIDM if the subhalos that host these galaxies, with $10~{\rm km~s^{-1}}\lesssim V_{\rm max}\lesssim 30~{\rm km~s^{-1}}$, are in the core-expansion phase for $\sigma_{\rm eff}/m\lesssim{\cal O}(10)~{\rm cm^2~g^{-1}}$~\citep{Valli171103502,Read180506934,Kaplinghat190404939,Silverman220310104}. This has motivated the exploration of SIDM scenarios with stronger DM self-interactions~\citep{Zavala190409998,Correa200702958,Turner201002924,Yang221113768}, such that some subhalos could be in the core collapse phase, resulting in a high density as observed in, e.g., Draco~\citep{Nishikawa190100499,Sameie190407872}. For our SIDM model, $\sigma_{\rm eff}/m\sim140~{\rm cm^2~g^{-1}}$ for the subhalos of the MW, such that many of them core collapse within the Hubble time.

For isolated spiral galaxies, SIDM models with $\sigma_{\rm eff}/m\sim3~{\rm cm^2~g^{-1}}$ can largely explain the diversity of galactic rotation curves over a wide mass range of $30~{\rm km~s^{-1}}\lesssim V_{\rm max}\lesssim 300~{\rm km~s^{-1}}$~\citep{Kamada161102716,Ren:2018jpt}. This cross section amplitude should be regarded as a lower limit for most of these systems, as the fit will not change significantly when the cross section increases further~\citep{Kaplinghat191100544}. Our SIDM model has $2~{\rm cm^2~g^{-1}}\lesssim \sigma_{\rm eff}/m\lesssim100\textup~{\rm cm^2~g^{-1}}$ over this $V_{\rm max}$ range, so we expect that the success of previous SIDM fits will largely remain, though a detailed reanalysis is needed. We will discuss the implications of our SIDM model for MW satellites and spiral galaxies further after presenting our results.

\section{Group Simulation} We present a pair of CDM and SIDM cosmological DM-only zoom-in simulations centered on a host halo with $M_{200}=1.1\times 10^{13}~\msun$ using high-resolution initial conditions from the ``Group'' suite of Symphony simulations \citep{Nadler220902675}; throughout, we measure masses that enclose $200$ times the critical density of the universe at $z=0$. These simulations are run with a high-resolution particle mass of $3\times 10^5~h^{-1}~\msun$ and a Plummer-equivalent gravitational softening length of $170~h^{-1}~\mathrm{pc}$ using \textsc{GADGET-2} \citep{Springel0505010}. Halos are identified with \textsc{ROCKSTAR}~\citep{Behroozi11104372} and merger trees are built using \textsc{consistent-trees}~\citep{Behroozi11104370}. The cosmological parameters are $\Omega_M=0.286$, $\Omega_\Lambda=0.714$, $n_s=0.96$, $h=0.7$, and $\sigma_8=0.82$ \citep{WMAP9}. 
DM self-interactions are implemented using the code developed and validated in~\cite{Yang220503392}, based on a velocity-dependent viscosity cross section; see Figure~\ref{fig:bmxs} (inset) and Appendix~\ref{sec:cross_section} for details.

\begin{figure*}[t!]
  \includegraphics[width=\textwidth]{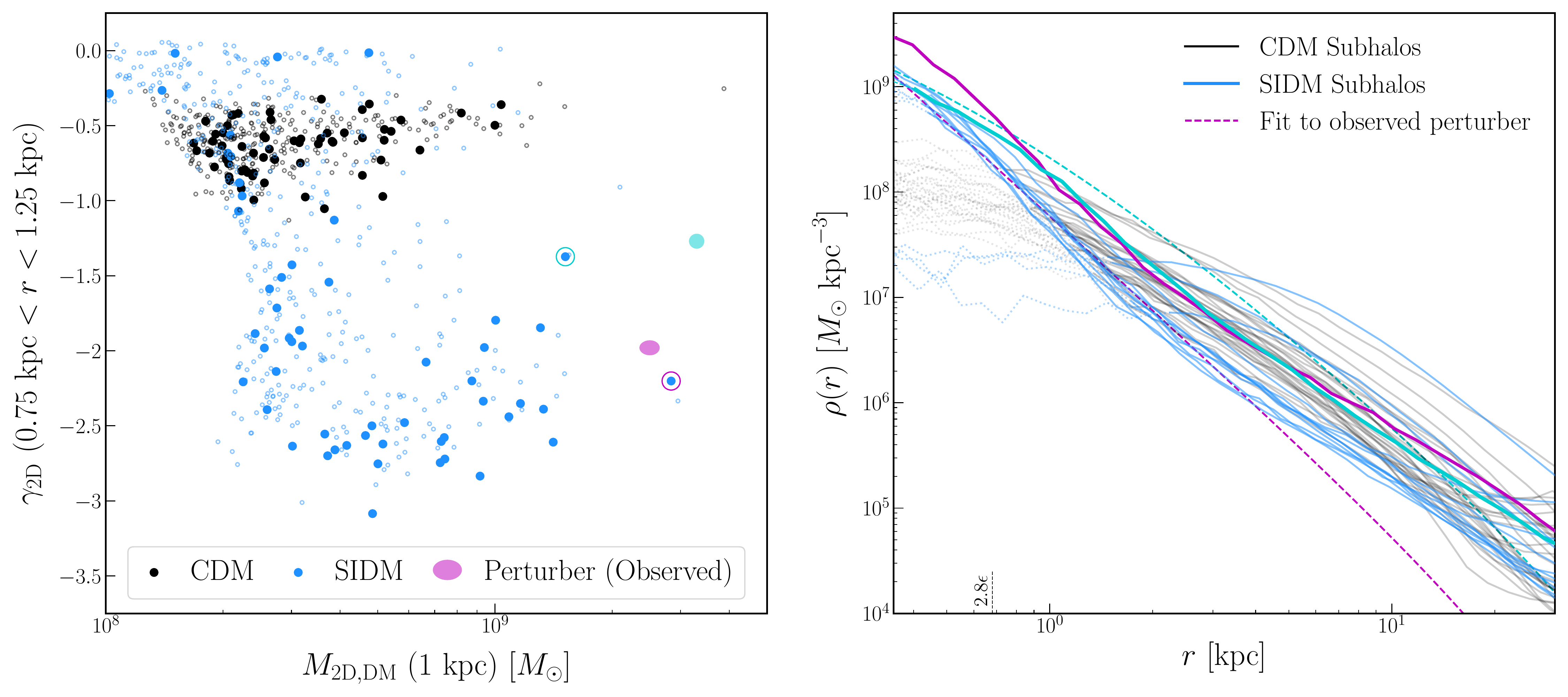}
  \caption{\label{fig:lens} Left panel: projected logarithmic density profile slope, averaged from $0.75$ to $1.25~\mathrm{kpc}$, versus enclosed projected mass within $1~\mathrm{kpc}$ for all subhalos of the group-mass host with $V_{\mathrm{max}}>30~\kms$ in our CDM (black) and SIDM (blue) simulations; isolated halos are shown by small unfilled points. The corresponding properties and $1\sigma$ uncertainties for the SDSSJ0946+1006 perturber derived in \citetalias{Minor201110627} are shown for a truncated NFW profile (tNFW; magenta) and when including higher-order multipoles in the lens model (tNFWmult; cyan). 
  Right panel: density profiles of subhalos with $V_{\mathrm{max}}>30~\kms$ in CDM (black) and SIDM (blue).  
  Lines become dotted when fewer than $1000$ particles are enclosed; the vertical dashed line shows a resolution limit of $2.8\epsilon = 680~\pc$. 
  Dashed magenta (cyan) lines show best-fit tNFW (tNFWmult) profiles from \citetalias{Minor201110627} for the observed perturber. The SIDM halo closest to the tNFW (tNFWmult) perturber is circled in magenta (cyan) in the left panel and shown by the solid magenta (cyan) line in the right panel; Appendix~\ref{sec:density_supplement} shows that the magenta density profile is measured smoothly.}
\end{figure*}

We study (sub)halos within $6~\mathrm{Mpc}$ of the host halo center or $\approx 10$ times the virial radius of the group-mass host. High-resolution particles comprise $>90\%$ of the mass in this $(6~\mpc)^3$ region, which is the effective volume of our zoom-in simulation. Furthermore, we only analyze (sub)halos that contain $>2000$ particles at $z=0$, corresponding to a present-day mass threshold of $8.6\times 10^8~\msun$; there are $\approx 10^3$ such isolated halos throughout the high-resolution region, several hundred of which are in the mass range relevant for the UDGs studied in \citetalias{Kong220405981}. Several large isolated halos are resolved in addition to the group-mass host, including a $\sim 10^{12}~\msun$ (MW-mass) halo and eight $\sim 10^{11}~\msun$ (Large Magellanic Cloud-mass) halos. Here, we focus on isolated halos and subhalos of the group-mass host, where subhalos are defined as halos whose centers are within the virial radius of the group host. Convergence tests and subhalo population analyses will be presented in a companion study (Nadler et al.\ 2024, in preparation).

At $z=0$, the group-mass host in our SIDM simulation has a $\approx 25~\mathrm{kpc}$ core, which is consistent with the core size expected semianalytically given our SIDM cross section~\citep{Kaplinghat13116524,Kaplinghat150803339,Robertson200907844}. Isolated halo mass functions are nearly identical in our CDM and SIDM simulations. However, as we will demonstrate, our Group SIDM model significantly diversifies isolated halos' and subhalos' inner density profiles, simultaneously yielding systems that resemble low-concentration UDG halos and the dense strong-lens perturber.\\

\section{Strong-lens Perturber (Sub)halos} The left panel of Figure~\ref{fig:lens} shows the inner projected mass and density profile slope for all subhalos of our group-mass host with $V_{\mathrm{max}}>30~\kms$, in our CDM (black) and SIDM (blue) simulations. We measure projected logarithmic density slopes averaged between $0.75$ and $1.25~\mathrm{kpc}$ following~\citetalias{Minor201110627}; within this range, our results are insensitive to the exact radii chosen for this averaging procedure. For comparison, we show the parameter space inferred by~\citetalias{Minor201110627} for the strong-lens perturber, assuming a truncated-NFW (``tNFW'') profile for the perturber and including higher-order multipoles in the lens model (``tNFWmult'').

Our CDM simulation yields subhalos with inner projected density profile slopes clustered around $-0.75$, consistent with the findings in~\citetalias{Minor201110627} based on IllustrisTNG~\citep{Nelson181205609}. Meanwhile, most surviving subhalos in our SIDM simulation have steeper inner density profiles than their CDM counterparts. Thus, the tension between the steepness of the strong-lens perturber's inner density profile and CDM predictions is significantly eased in our SIDM model. Only one of our SIDM subhalos approaches the required inner mass of $2\times 10^9~\msun$; however, our group host is on the low end of the halo mass distribution for SDSSJ0946+1006, between $1$ and $6\times 10^{13}~\msun$ (\citealt{Auger10072880}; \citetalias{Minor201110627}). A modest increase in host halo mass is therefore likely to produce even more subhalos with enclosed masses similar to the SDSSJ0946+1006 perturber in our SIDM model. We highlight two SIDM subhalos that match the properties of the strong-lens perturber fairly well; both of these systems have high concentrations and early formation times relative to the rest of the SIDM subhalo population, and one of them undergoes a pericentric passage $\approx 4~\Gyr$ ago, which may accelerate its gravothermal evolution~\citep{Nishikawa190100499,Sameie190407872}.

We also find a population of \emph{isolated} SIDM halos with properties similar to the SDSSJ0946+1006 perturber, shown as unfilled points in the left panel of Figure~\ref{fig:lens}. \citetalias{Minor201110627} found that essentially no isolated halos in IllustrisTNG have sufficiently steep inner density slopes to explain the perturber's properties. Thus, our SIDM model also eases the $\sim 5\sigma$ tension between CDM predictions and the SDSSJ0946+1006 perturber when interpreted as a line-of-sight halo. This is intriguing as~\cite{Sengul:2021lxe} and \cite{Sengul:2022edu} reported that another lens, JVAS B1938+666~\citep{King:1997tr,Vegetti12013643}, is perturbed by a dense line-of-sight structure.

The right panel of Figure~\ref{fig:lens} compares subhalo density profiles from our CDM and SIDM simulations to that inferred for the observed perturber assuming a tNFW (magenta) or tNFWmult (cyan) profile. At the $1~\mathrm{kpc}$ radius where the perturber profile is best constrained (corresponding to $r/R_{\mathrm{vir}}=0.025$ in Figure~\ref{fig:lens}), our SIDM predictions span the inferred density profiles, which are most robustly constrained within $\approx 1~\kpc$, indicating a remarkable level of agreement that is difficult to achieve in CDM.

\begin{figure*}[t!]
  \centering
  \includegraphics[width=\textwidth]{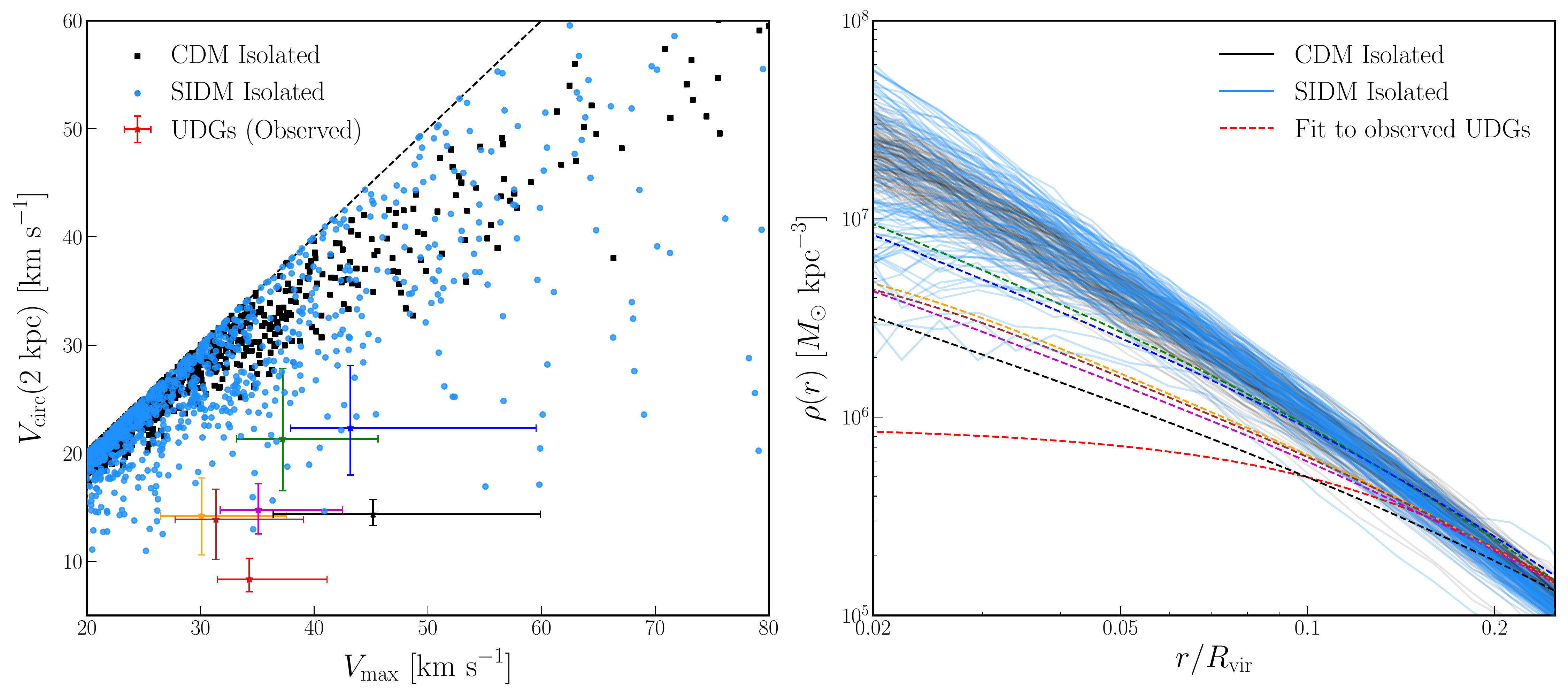}
  \caption{\label{fig:udg} Left panel: circular velocity evaluated at $2~\mathrm{kpc}$ vs.\ maximum circular velocity for seven isolated UDGs studied in \citetalias{Kong220405981}, with means and $1\sigma$ uncertainties shown for AGC 114905 (red), 122966 (green), 219523 (blue), 248945 (orange), 334315 (magenta), 749290 (brown), and 242019 (black). Black (blue) points show the same quantities for isolated halos throughout the $6~\mathrm{Mpc}$ high-resolution region of our CDM (SIDM) simulation. Only halos resolved with more than $2000$ particles at $z=0$ are shown; the dashed line shows the one-to-one relation.  
  Right panel: density profiles of isolated halos with $30~\kms<V_{\mathrm{max}}<50~\kms$ and $M_{200}>5.5\times 10^9~\msun$ in our CDM (black) and SIDM (blue) simulations. 
  Lines become dotted when fewer than $1000$ particles are enclosed. 
  Dashed lines show best-fit \cite{Read:2015sta} density profiles from \citetalias{Kong220405981} for each observed UDG.}
\end{figure*}

Most surviving subhalos in our SIDM simulation have higher-amplitude and steeper density profiles than their CDM counterparts in the same $V_{\mathrm{max}}$ range. 
By studying the evolution of matched CDM and SIDM subhalos, we find that initially high-mass subhalos with large cores at infall are quickly disrupted in our SIDM simulation, mainly due to tidal stripping \citep{Vogelsberger12015892,Dooley160308919,Nadler200108754,Yang210202375,Slone:2021nqd,Zeng211000259}. Meanwhile, subhalos with relatively low masses and high concentrations at infall often show signs of core collapse after their first pericenter, consistent with findings that tidal stripping accelerates subhalos' gravothermal evolution \citep{Nishikawa190100499,Sameie190407872,Kahlhoefer190410539,Correa200702958,Zeng211000259}. These core-collapsed subhalos are more resilient to disruption than cored systems; thus, most surviving subhalos in our Group SIDM model have steep inner density profiles.

\section{UDG Halos} The left panel of Figure~\ref{fig:udg} shows halo circular velocity, evaluated at a radius of $2~\mathrm{kpc}$, versus maximum circular velocity for the seven UDGs from~\citetalias{Kong220405981} based on the kinematic data presented in~\cite{ManceraPina:2019zih,ManceraPina:2020ujo,PinaMancera211200017} and \cite{Shi:2021tyg} for AGC 114905 (red), 122966 (green), 219523 (blue), 248945 (orange), 334315 (magenta), 749290 (brown), and 242019 (black). The same quantities for isolated halos in our CDM (SIDM) simulation are shown by black (blue) points. We find that isolated halos with $V_{\mathrm{circ}}(2~\mathrm{kpc})\lesssim 20~\kms$ and $30~\kms<V_{\mathrm{max}}<50~\kms$ are extremely rare in CDM, for which no halos in our sample fall within the $1\sigma$ uncertainties of the observed UDGs' $V_{\mathrm{circ}}(2~\mathrm{kpc})$ values in the relevant $V_{\mathrm{max}}$ range. Given the small volume of our high-resolution region, which only yields $\approx 10^3$ isolated halos in the UDG mass range, our result is consistent with the findings in \citetalias{Kong220405981} that (i) UDG-like halos are $\gtrsim 3\sigma$ to $5\sigma$ away from the median CDM expectation in IllustrisTNG, and (ii) even extremely low-concentration CDM halos’ inner profiles are inconsistent with the circular velocities of the observed UDGs.

Meanwhile, our Group SIDM model diversifies the $V_{\mathrm{circ}}(2~\mathrm{kpc})$--$V_{\mathrm{max}}$ relation and produces a population of halos that is consistent with UDGs' inner circular velocities. Consistent with \citetalias{Kong220405981}, we find that these UDG analog halos preferentially form later and have higher spin parameters than the rest of the isolated SIDM halo population; thus, a detailed study of our UDG analogs' secondary halo properties is a promising area for future work.

Next, the right panel of Figure~\ref{fig:udg} shows the density profiles of all isolated halos with $30~\kms<V_{\mathrm{max}}<50~\kms$ and $M_{200}>5.5\times 10^9~\msun$ in our CDM and SIDM simulations. This $V_{\mathrm{max}}$ range encompasses the observed UDGs, and the mass cut ensures that all UDGs' halo masses are larger than the universal DM-to-baryon ratio using the $1\sigma$ lower limits from~\citetalias{Kong220405981}. Roughly half of the resulting halos in our SIDM simulation have kiloparsec-scale cores. Dashed lines show best-fit \cite{Read:2015sta} density profiles of the observed UDGs from~\citetalias{Kong220405981}; our SIDM predictions span the UDGs' inferred profiles, and this level of agreement is again difficult to achieve in CDM.

We note that there are detailed differences between our simulated SIDM profiles and the UDG profiles from~\citetalias{Kong220405981}. Among the UDGs, AGC 114905~\citep{PinaMancera211200017} has the lowest inner halo density and the largest core size; our SIDM halos are still too dense in the inner regions to fit this system. It is possible that AGC 114905's halo concentration is even lower than any halos in our simulation. To test this, we used a parametric model to predict SIDM profiles as a function of halo  concentration~\citep{Yang:2023jwn}; we found that our SIDM model can fit AGC 114905 for $c_{200}\approx2.2$ with $M_{200}\approx1.4\times10^{10}~{M_\odot}$ ($V_{\rm max}\approx34~{\rm km~s^{-1}}$ and $\sigma_{\rm eff}/m\approx99~{\rm cm^2~g^{-1}}$). This concentration would be ``$7\sigma$" below the median according to the standard $c_{200}\textup{--}M_{200}$ relation~\citep{2014MNRAS.441.3359D}. However, the low-concentration tail of this relation is a power law, rather than a Gaussian. Thus, the probability of such a low-concentration halo is small but not negligible as found in the IllustrisTNG simulation~\citetalias{Kong220405981}, which has a much larger volume than our zoom-in.

For the remaining UDGs, our simulated SIDM halos are slightly more cored than the observed systems~\citetalias{Kong220405981}. In fact, these UDGs have rather sharply rising rotation curves and they do not strongly prefer a cored density profile, although their halos must have a low concentration. It is important to investigate whether the trend holds with more kinematic data (the fits in~\citetalias{Kong220405981} were based on two data points for each rotation curve, except for AGC 114905 and 242019). Furthermore, the presence of gas, which is dynamically important, may also bring our predicted SIDM density profiles into better agreement with observations.

A more detailed comparison between simulated and observed systems will require modeling the baryonic contribution and potentially expanding the \cite{Read:2015sta} density profile to encompass our diverse SIDM halos. It would also be interesting to explore the formation of gas-rich galaxies in our Group SIDM model using hydrodynamic simulations, since the gas distribution could be extended due to the presence of large DM cores, potentially providing a novel explanation of UDGs' observed sizes. Note that the majority of cored field halos in our SIDM simulation are isolated throughout their histories, differentiating our model from scenarios in which environmental effects like tidal stripping produce shallow CDM density profiles (e.g., see~\citealt{Jiang:2018iut,Benavides210901677,Benavides220907539}).

In addition to core-forming halos, we also find a sizable population of isolated SIDM halos with higher-amplitude inner density profiles and smaller values of $R_{\mathrm{max}}$ than any halos in our CDM simulation (where $R_{\mathrm{max}}$ is the radius at which $V_{\mathrm{max}}$ is achieved), indicating efficient core collapse in our SIDM model. These core-collapsed halos' inner logarithmic density profile slopes are often as steep as $-3$; they appear near the one-to-one relation in Figure~\ref{fig:udg}~\citep{Yang221113768}.\footnote{Some deeply core-collapsed halos' rotation curves are declining at $2~\mathrm{kpc}$, but the vast majority of such systems lie below the mass cut in Figure~\ref{fig:udg}.} The fraction of isolated core-collapsed halos in our SIDM simulation is consistent with analytic estimates of $t_c$ in our SIDM model, which predict that higher-concentration halos collapse more quickly~\citep{Essig180901144}.

Because our SIDM scenario diversifies isolated halos \emph{in both directions} relative to CDM, it predicts that extremely dense, isolated galaxies with rapidly rising rotation curves should be observed in addition to the UDGs. Rapidly rising rotation curves are particularly informative if the baryonic contribution to the rotation curve is small; existing data sets may hint at the existence of such galaxies~\citep{Santos-Santos191109116,deNaray09123518}, motivating detailed comparisons to simulations like ours.

\section{Conclusions} We have focused on two opposite extremes of low-mass systems, and their corresponding halo properties---a dense strong-lens perturber and DM-deficient gas-rich UDGs. Using the first group-scale simulation with large-amplitude, velocity-dependent DM self-interactions, we showed that SIDM can simultaneously produce halos at both extremes.

Our study reveals exciting directions for the study of strong lensing and UDG (sub)halos. It is timely to revisit predictions for the diversity of galactic rotation curves, which have mainly been studied for SIDM halos in the core-expansion phase~\citep{Kamada161102716,Ren:2018jpt,Creasey:2016jaq,Zentner220200012}, in the context of strong DM self-interactions. In our SIDM scenario, isolated halos are found in both core-expansion and core-collapse phases, yielding an even larger diversity of inner halo profiles than less extreme SIDM models. 
In parallel, this physics should be detectable in future strong lensing studies using both gravitational imaging and flux-ratio statistics (e.g., see \citealt{Minor:2016jou,Meneghetti200904471,Meneghetti230905799,Gilman210505259,Gilman220713111,Yang210202375,Loudas220913393,Sengul:2022edu,Dhanasingham230610109,Zhang230809739}); observations of MW satellites will also probe our SIDM model, which predicts that a substantial fraction of the low-mass subhalos that host faint dwarf galaxies are core collapsed (e.g., \citealt{Kaplinghat190404939,Correa200702958, Turner201002924,Silverman220310104,Slone:2021nqd,Yang221113768}).

Exploring a range of strong, velocity-dependent cross sections will be necessary to robustly constrain our SIDM scenario. We expect that models with $\sigma_0/m$ smaller by an ${\cal O}(1)$ factor, keeping $w\sim100~{\rm km~s^{-1}}$, can also produce halos at both density extremes. Thus, observational constraints on the fraction of core-collapsed halos will narrow the favored SIDM parameter space. Such constraints will be enabled by combining (semi)analytic predictions for SIDM halo populations~\citep{Jiang220612425,Shah230816342,Yang:2023jwn,Yang:2023stn}, galaxy--halo connection modeling techniques~\citep{Nadler180905542}, and observations of DM structure and galaxies over a wide mass range.\\ \\ \\


\begin{acknowledgments} 

We thank Manoj Kaplinghat for helpful discussions, Pavel Mancera Pi\~{n}a for comments on the manuscript, and Demao Kong for providing data from \citetalias{Kong220405981}. This work was supported by the John Templeton Foundation under grant ID \#61884 and the U.S. Department of Energy under grant No.\ de-sc0008541 (D.Y.\ and H.-B.Y.). The opinions expressed in this publication are those of the authors and do not necessarily reflect the views of the John Templeton Foundation. This work was performed in part at the Aspen Center for Physics, which is supported by National Science Foundation grant PHY-2210452. The computations presented here were conducted through Carnegie's partnership in the Resnick HPC Center, a facility supported by Resnick Sustainability Institute at the California Institute of Technology.

\end{acknowledgments}

\bibliographystyle{yahapj2}
\bibliography{reference}


\clearpage

\onecolumngrid
\appendix

\section{SIDM Cross Section Fit}
\label{sec:cross_section}

We assume Rutherford-like scattering in the Born limit~\citep{Feng09110422,Ibe:2009mk,Tulin13023898} and perform the SIDM simulation using the viscosity cross section defined in \cite{Yang220503392}, i.e.,
\begin{align}
\label{eq:sigmav}
\sigma_{V}(v)&=\frac{3}{2}\int \mathrm{d}\cos\theta\sin^2\theta\frac{\mathrm{d}\sigma}{\mathrm{d}\cos\theta}& \nonumber \\ &=\frac{6\sigma_0 w^6}{v^6}\left[\left(2+\frac{v^2}{w^2}\right)\ln\left(1+\frac{v^2}{w^2}\right)-\frac{2v^2}{w^2}\right],&
\end{align}
where we take $\sigma_0/m=147.1~{\rm cm^2~g^{-1}}$ to be the same as that used in the MW analog simulation from \cite{Yang221113768}, while setting $w=120~{\rm km~s^{-1}}$. Considering a scenario where DM particles interact with a dark photon, we have $\sigma_0=4\pi\alpha^2_{\chi}/(m^2 w^4)$ and $w=m_\phi c/m$, where $\alpha_\chi$ is the fine structure constant in the dark sector, $m_\phi$ is the mass of the dark photon, and $c$ is the speed of light. \cite{Yang220503392} showed that the viscosity cross section provides an excellent approximation for modeling angular- and velocity-dependent DM self-scattering. If DM is made of one particle species, the viscosity cross section for M\o{}ller scattering would be more appropriate~\citep{Yang220503392,Girmohanta:2022dog}, as it includes both $t$- and $u$-channel contributions, as well their interference term. However, the difference is small and we can reinterpret the cross section of Rutherford scattering to that of M\o{}ller scattering by rescaling particle parameters within less than $10\%$~\citep{Girmohanta:2022izb}. In addition, our simulation assumes all particles participate in scattering; see \cite{Yang220503392} for a self-consistent interpretation in terms of quantum statistics.

\begin{figure}[t!]
  \centering
  \includegraphics[width=0.485\textwidth]{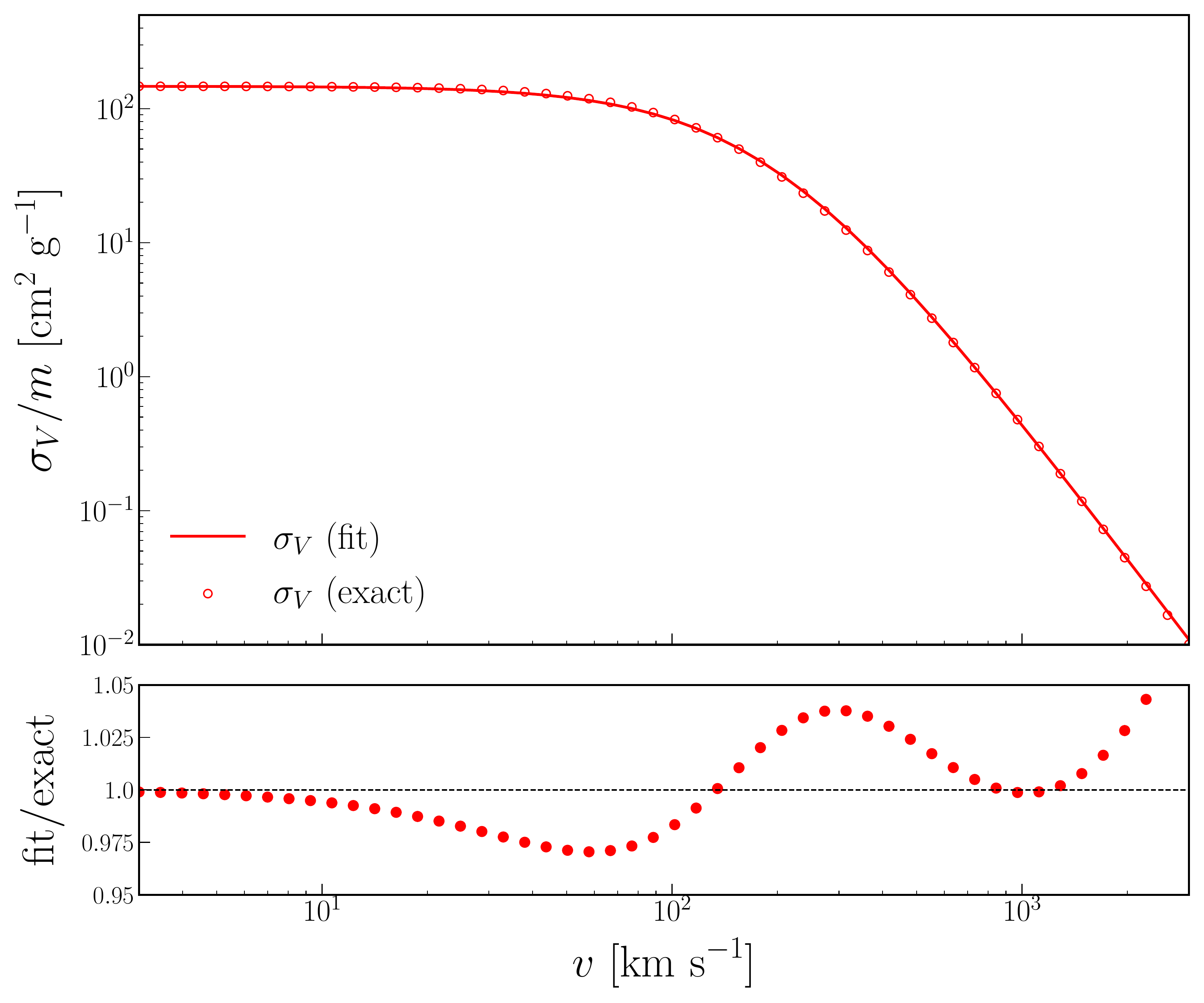}
  \caption{\label{fig:xsec} Top panel: the exact viscosity cross section for our Group SIDM model Equation~\ref{eq:sigmav} (red line) and the fitting function implemented in our simulation Equation~\ref{eq:fit} (unfilled red circles). Bottom panel: ratio of the fitting function to the exact viscosity cross section, illustrating that they agree at the percent level.
 }
\end{figure}

For simplicity, we fit the exact viscosity cross section Equation~\ref{eq:sigmav} with the function
\begin{equation}
\label{eq:fit}
    \sigma_{V,\mathrm{fit}}(v) = \frac{\sigma_0}{\left[1+\left(v/v_0\right)^{\delta}\right]^{\beta}},
\end{equation}
where $\sigma_0/m=147.1~\mathrm{cm}^2~\mathrm{g}^{-1}$, $v_0=190~\mathrm{km\ s}^{-1}$, $\delta=1.72$, and $\beta=2$. Figure~\ref{fig:xsec} shows that this fit matches the viscosity cross section for our Group SIDM model at the percent level over the relative velocity range of interest. Thus, our implementation accurately captures the underlying SIDM cross section.

\section{Density Profile of the Simulated Perturber Analog}
\label{sec:density_supplement}

We compute surface mass density by integrating the 3D density profile along an arbitrary $z$-axis as 
\begin{equation}
    \Sigma (R) = \int_{-R_{\mathrm{vir}}}^{R_{\mathrm{vir}}} \rho(\sqrt{R^2+z^2})~\mathrm{d}z,\label{eq:sigma_int}
\end{equation}
assuming that the halo is spherically symmetric. Accordingly, the 2D mass enclosed within $1$ kpc is computed as
\begin{equation}
M_{\rm 2D} (1~\mathrm{kpc}) = 2\pi \int_0^{1~\text{kpc}} \Sigma(R) R~\mathrm{d}R. 
\end{equation}

\begin{figure*}[t!]
 \centering
 \includegraphics[height=0.3\textwidth] {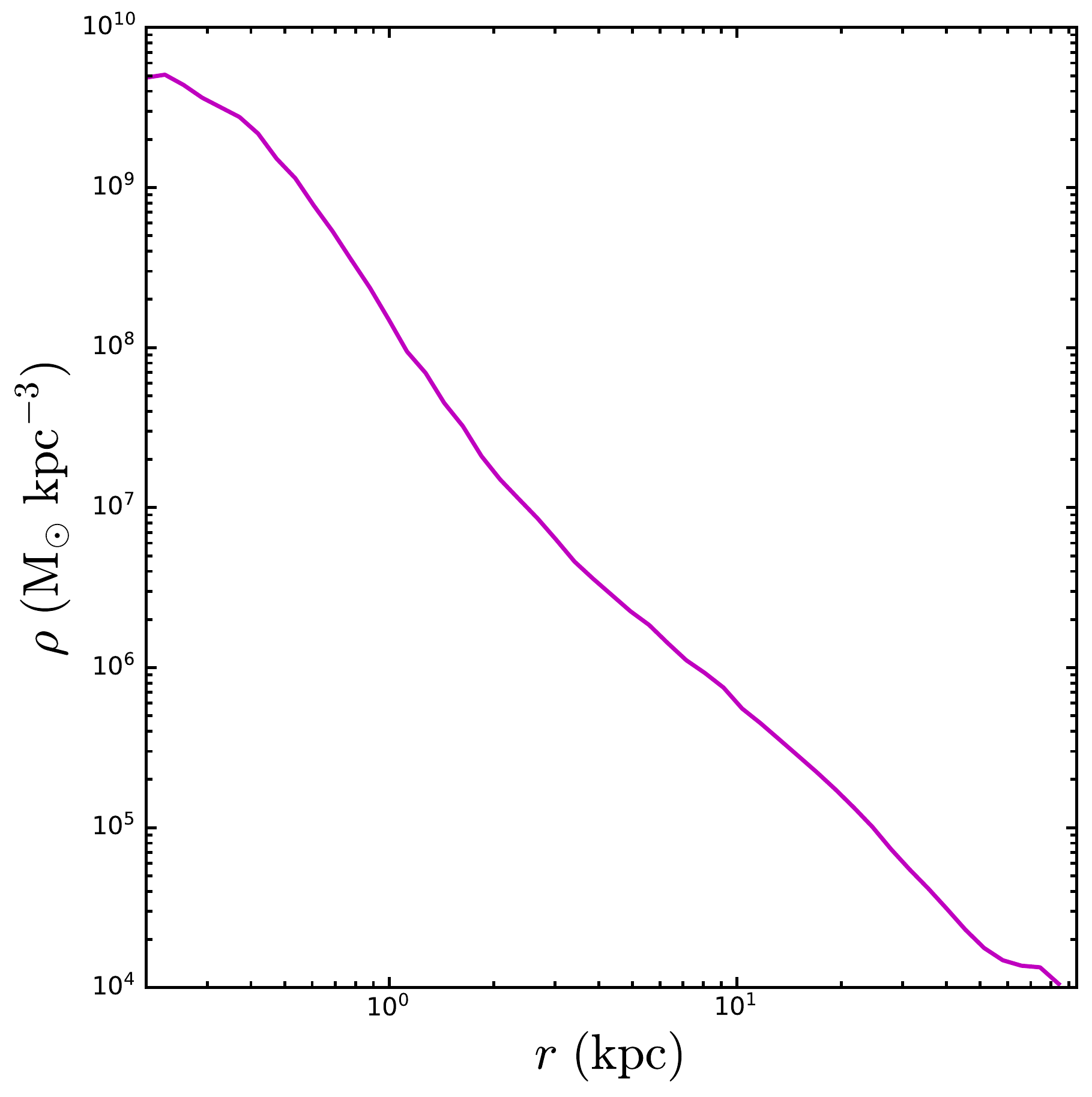}
 \includegraphics[height=0.3\textwidth] {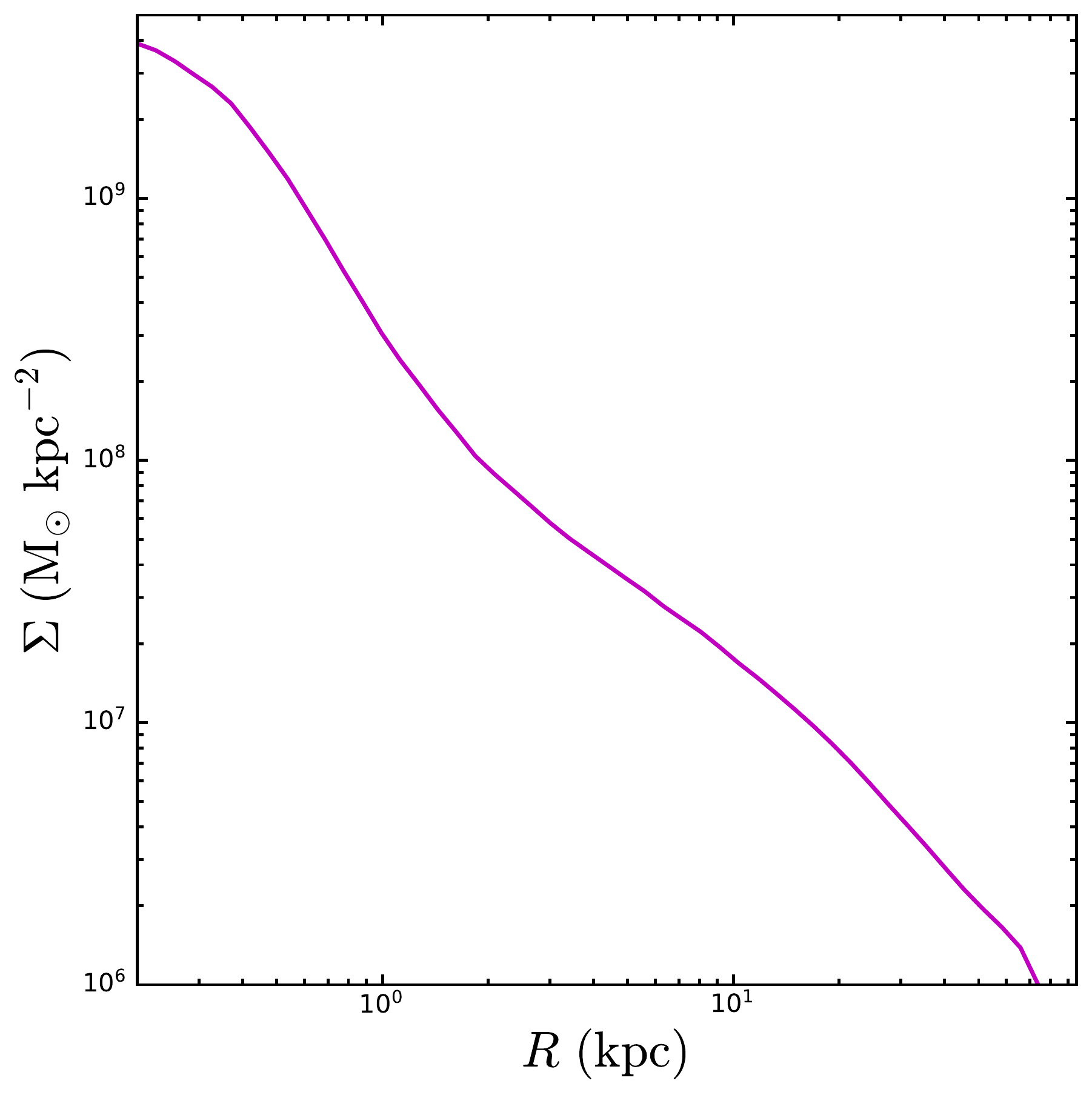}
 \includegraphics[height=0.3\textwidth] {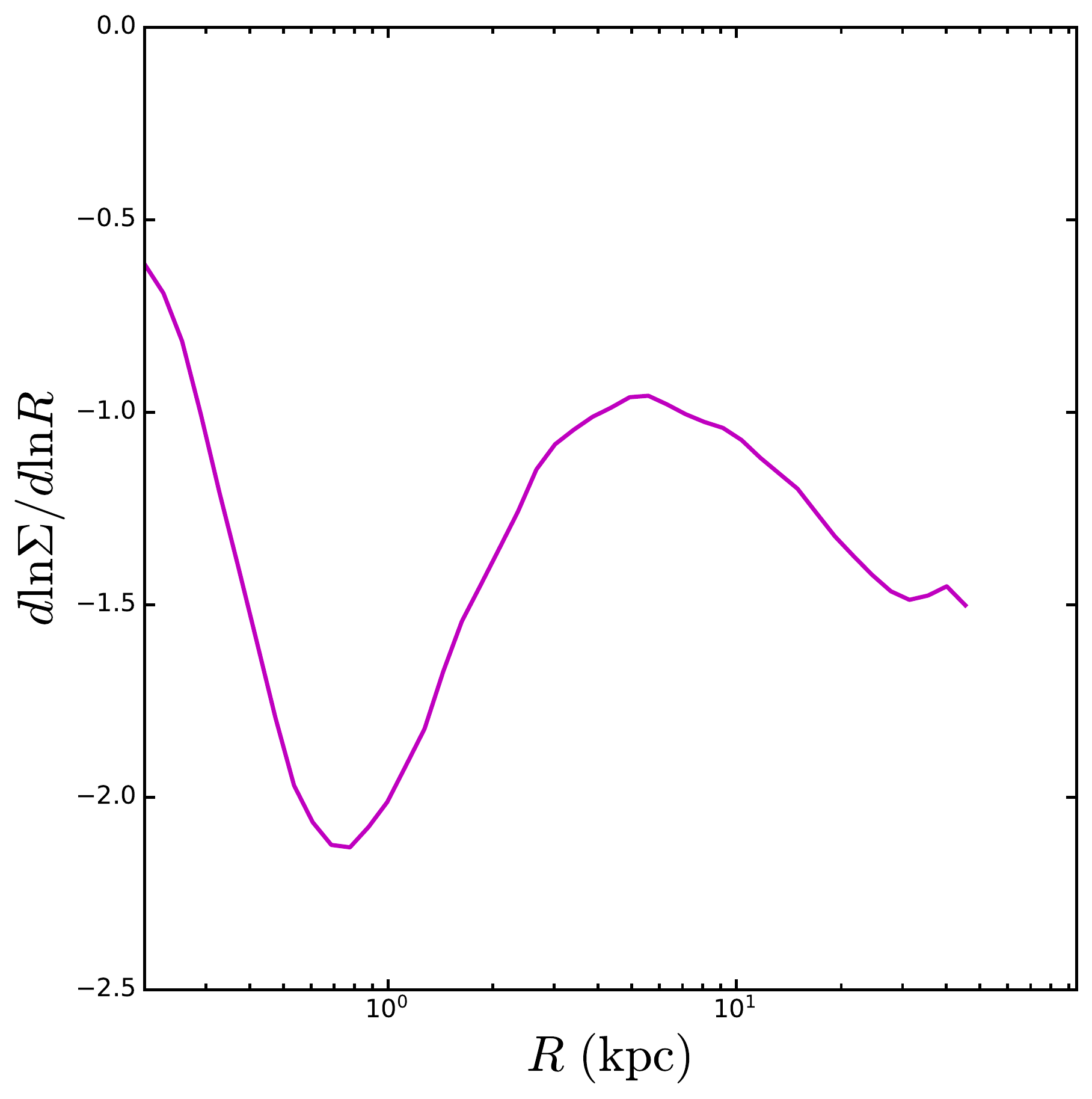}
  \caption{\label{fig:subpert} The 3D density profile (left), surface density (middle), and project density slope profile for the tNFW strong-lensing perturber analog from our SIDM simulation, corresponding to the magenta halo in Figure~\ref{fig:lens}.}
\end{figure*}

We construct an interpolation function to smoothly model surface density versus projected radius by evaluating this relation on a log--log grid.
The slope $\gamma_{\rm 2D}\equiv \mathrm{d}\ln\Sigma/\mathrm{d}\ln R$ is then numerically extracted from the interpolation function as a function of $\ln R$. 
To suppress numerical uncertainties, we present averaged $\gamma_{\rm 2D}$ between $0.75$ and $1.25$ kpc, following \citetalias{Minor201110627}. Figure~\ref{fig:subpert} shows the 3D density profile (left), surface density profile (middle), and projected density profile slope for the tNFW strong-lensing perturber analog from our SIDM simulation (corresponding to the magenta density profile in the right panel of Figure~\ref{fig:lens}). All three profiles are measured smoothly down to scales comparable to our simulation's softening length, and the results in the inner regions are robust to changes in the limits of integration in Equation.~\ref{eq:sigma_int}.

\end{document}